\documentclass[useAMS,usenatbib,usegraphicx]{mn2e}

\def\kms{km s$^{-1}$}
\def\hi{H{\sc i}}
\def\hii{H{\sc ii}}
\def\Msun{M$_{\odot}$}

\def\cmdos{cm$^{-2}$}
\def\cm{cm$^{-3}$}

\def\gra{$^{\circ}$}

\title[The ISM linked to Sh2-132]{Ionized gas, molecules, and dust in Sh2-132}

\author[Vasquez et al.]{J. Vasquez$^{1,2}$\thanks{E-mail:
jvasquez@fcaglp.unlp.edu.ar; pete@iar-conicet.gov.ar}, 
C.E. Cappa$^{1,2}$, S. Pineault$^{1,3}$ and N.U. Duronea$^{1,2}$ \\
$^{1}$Instituto Argentino de Radioastronomia, CCT-La Plata, CONICET, 
C.C.5., 1894, Villa Elisa, Argentina\\
$^{2}$Facultad de Ciencias Astron\'omicas y Geof\'{\i}sicas, Universidad 
Nacional de La Plata, La Plata, Argentina\\
$^{3}$D\'epartment de physique, de g\'enie physique et d'optique and 
Centre de recherche en astrophysique du Qu\'ebec (CRAQ), \\
Universit\'e Laval, Qu\'ebec, Canada GIVOA6}

\begin{document}

\date{Accepted 1988 December 15. Received 1988 December 14; in original form 1988 October 11}

\pagerange{\pageref{firstpage}--\pageref{lastpage}} \pubyear{2002}

\maketitle

\label{firstpage}

\begin{abstract}

We analyze the various interstellar components of the \hii\ region Sh2-132. 
The main stellar source is the double binary system that includes the 
Wolf-Rayet star WR153ab. We use radio continuum images at 408 and 1420 MHz, 
and \hi\ 21cm line data taken from the Canadian Galactic Plane Survey, 
molecular 
observations of the $^{12}$CO(1-0) line at 115 GHz from  the Five College 
Radio Astronomy Observatory,  and available mid and far IR observations 
obtained with the MSX and IRAS satellites, respectively.
 
Sh2-132 is composed of two shells showing radio continuum counterparts at 
both frequencies. The emission is thermal in nature. The estimated rms 
electron density and ionized mass of the nebula  are 
$n_e \simeq 20\,\rm cm^{-3}$ and $M_{\rm HII}\simeq 1500 M_{\odot}$. 
The distribution of the CO emission shows molecular gas bordering the
ionized nebula and interacting with it. The velocities of the molecular
gas is in the range --38 to --53 \kms, similar to the velocity of the
ionized gas.
 
The emission at 8.3 $\mu$m reveals a ring like feature of about 15\arcmin\
that encircles the bright optical regions. This emission is due to the
PAHs and marks the location of photodissociation regions.

The gas distribution in the environs of Sh2-132 can be explained in a
scenario where the massive stars in the region 
photodissociated, ionized, and swept-up the dense molecular material
from the parental cloud through their strong stellar winds and intense
UV photon flux. 

\end{abstract}


\section{Introduction}\label{intro}

The interaction between massive stars and the interstellar medium leads
to the formation of \hii\ regions, interstellar bubbles, and new stars. 
More specifically, the interaction between the molecular medium and the UV 
stellar radiation is at the origin of photodissociation regions (PDRs).
Molecules are dissociated by UV photons in the range 6 eV to 13.6 eV.
The gas component ranges from ionized atoms at the PDR surface to molecules 
deeper into the molecular region as the FUV flux is absorbed. The FUV flux 
illuminating these regions  range between 1$\le G_0 \le$ 10$^6$, where 
$G_0$ is the intensity of the radiation field in the solar neighborhood 
(Habing 1968), while the \hi\ density lies between 10 \cm\ 
$\le n_H \le$ 10$^5$ \cm\ (Tielens \& Hollembach 1985, Sternberg \& Dalgarno 
1989, Sternberg \& Dalgarno 1995, Bakes \& Tielens 1998). PDRs are
formed at the edge of \hii\ regions, planetary nebulae, reflexion nebulae,
and stellar forming regions. Forbidden spectral lines in the far infrared as
well as molecular lines are observed towards PDRs (e.g. Hogerheijde et al. 
1995, Kaufman et al. 1999).

An important component of PDRs is interstellar dust. Its presence plays a 
major role absorbing stellar radiation and reemitting it in the mid and far 
infrared. Polycyclic aromatic hydrocarbons (PAHs), which emit at 3.3, 6.2, 
7.7, 8.6, and 11.3 $\mu$m, are tracers of PDRs (e.g. Watson et al. 2008). 

In this paper, we study the interstellar environment of the \hii\ region 
Sh2-132 (Sharpless 1959), located in the Perseus spiral arm, 
paying particular 
attention to the presence of  PDRs at the interfase between the
ionized and molecular gas.

Sh2-132 was detected in H$\alpha$, [O III], and [S II] lines by Heckathorn 
et al. (1982). The nebula is centered approximately at $(l,b)\simeq$ 
(102\gra 50\arcmin, --0\gra 42\arcmin) and presents two bright regions, 
which are surrounded by diffuse emission regions. They can be identified
in Fig. 1, which displays the DSS-R image of the nebula.

Churchwell \& Walmsley (1973) performed radio continuum observations at 2695 
MHz with an angular resolution of 18\farcm 2, and, using measurements at
several frequencies,  derived a thermal spectral index. Harten et al. (1978) 
investigated the characteristics and morphology of the nebula based on 
radio continuum observations at 610 MHz using the WSRT with a synthesized 
beam of 60\arcsec. Most of the emission at this frequency originates from 
two optically bright regions named Shell A and Shell B, located at 
$(l,b)\simeq$ (102\gra 57\arcmin, \hbox{--0\gra 42\arcmin)} and $(l,b)\simeq$ 
(102\gra 47\arcmin, --0\gra 42\arcmin). Both shells coincide with the
bright optical regions within the nebula and are indicated  in Fig. 1.

According to Harten et al. (1978), the optically bright section of 
Sh2-132 (at $l>$102\gra 43\arcmin) is an ordinary \hii\ region excited by 
the massive O-type star BD+55\gra 2722 and the  WR star WR\,153ab 
(= HD\,211853), which are located close to the center of the nebula and 
linked to Shell B.    
Chu et al. (1983) reached to similar conclusions, based on 
the presence of the stars near the center of the nebula and on the absence 
of velocity changes in the nebular spectra. The main parameters of these 
stars, as well as of other OB stars listed by Harten et al. (1978) that appear 
projected onto the \hii\ region, are summarized in 
Table 1, which lists their galactic coordinates, 
spectral classification, visual magnitude $V$, colour indices ($B-V$) and 
($B-V$)$_0$, visual absorption $A_V$, absolute magnitude $M_V$, and 
derived spectrophotometric distance $d$. 
Uncertainties in the derived spectrophotometric distances  of some of the 
stars were estimated using  $M_V$ values from Vacca et al. (1996) and 
Landolt-B$\ddot{o}$rnstein (1982). The large distace derive for 
LS\,III+55\gra 37 casts doubts on their relation to the nebula.
Note that, based on photometric data, 
Panov \& Seggewiss (1990) concluded that HD\,211853 is a system with two 
pairs of stars, both hosting a WR component and an O-type star.

\begin{figure}
\includegraphics[width=84mm]{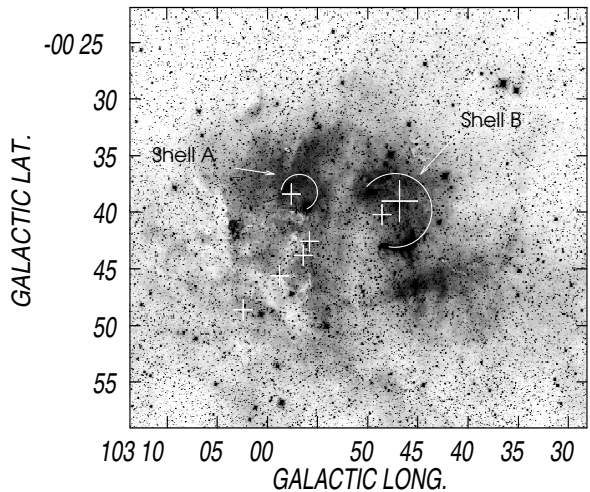}
\caption{DSS-R image of the brightest section of Sh2-132. The grey scale is 
arbitrary. The large cross marks the position of the WR star, while smaller
crosses indicate the location of the OB stars.}
\label{opt1-sh2-132}
\end{figure}

Chu \& Treffers (1981), based on Fabry Perot observations of the H$\alpha$
line, found that the ionized gas has LSR velocities in the range --53 
to --45 \kms, in agreement with previous results from Georgelin \& Georgelin 
(1976, --50.9 \kms), Reynolds (1988, --48$\pm$1 \kms), and Fich et al. 
(1990, --47.1 \kms). Quireza et al. (2006) detected ionized gas at --50.46 
and --49.48 \kms\ from He and H radio recombination lines, respectively. 
Harten et al. (1978) estimated an emission measure $EM$ = 3000 pc cm$^{-6}$
for the entire nebula.

\begin{table*}
\begin{center}
\caption{O-B type stars towards the bright section of Sh2-132}
\label{tabla-estrellas}
\begin{tabular}{lcccccccc}
\hline
Star       &$(l,b)$        & ST                 & $V$   & $(B-V)$ & $(B-V)_0$ &  $A_v$ & $M
_V$ &  $d$ \\
           &               &                    & mag   & mag     & mag       &mag         
  &mag    &   kpc   \\
\hline\hline
HD\,211853 & 102\gra 46\farcm 8,--0\gra 39\arcmin & WN6/WC+O6I$^{(1)}$ &  9.08$^{(2)}$ &  0.27$^{(2)}$ &   $^{(2)}$  & 2.28$^{(2)}$ & --6.4$^{(2)}$ & 2.75$^{(2)}$,5.0$^{(4)}$ \\
(WR\,153ab)\\
LS\,III+55\gra 37 &  102\gra 46\farcm 8,--0\gra 40\farcm 2 & O7V$^{(3)}$ & 11.01 &  0.32$^{(3)}$& --0.29$^{(6)}$  & 1.86 &  --5.2$^{(5)}$& 7.5   \\
BD+55\gra 2722   & 102\gra 48\farcm 6,--0\gra 40\farcm 2  & O8.5V$^{(3)}$  &  9.91$^{(3)}$   & 0.45$^{(3)}$ & --0.28$^{(6)}$ & 2.26 & --4.7${(5)}$ & 3.0  \\
LS\,III+55\gra 40     & 102\gra 56\farcm 4,--0\gra 42\farcm 6  & O9V$^{(3)}$ & 12.08$^{(3)}$   &  0.64$^{(3)}$  & --0.28$^{(6)}$  & 2.85 & --4.5${(5)}$ & 5.6 \\
LS\,III+55\gra 42     & 102\gra 56\farcm 4,--0\gra 43\farcm 8  & B$^{(7)}$   & 12.38   & 0.52 & --0.26$^{(6)}$  & 2.42 & --2.45, --5.1$^{(*5)}$ & 5-0-10.3 \\
LS\,III+55\gra 39     & 102\gra 57\farcm 6,--0\gra 38\farcm 4  & B$^{(7)}$ & 11.74 &  0.45 & --0.26$^{(6)}$ & 2.20 & --2.45, --5.1$^{(*5)}$ &  2.5-8.5 \\
LS\,III+55\gra 45 & 102\gra 59\farcm 4,--0\gra 45\farcm 6 & O8$^{(3)}$  & 10.45$^{(3)}$ &  0.70$^{(3)}$  & -0.27$^{(5)}$ & 3.01 & --4.9, --5.8$^{(**5)}$ & 2.9-4.9 \\
LS\,III+55\gra 49      & 103\gra 02\farcm 4,--0\gra 48\farcm 6 & B$^{(7)}$ & 11.50& 0.57 & --0.26$^{(6)}$  & 2.5 7 & --2.45, --5.1$^{(*5)}$ & 1.9-6.4 \\
\hline
\end{tabular}
\\
\footnotesize{References: (1) Smith et al. (1998), (2) van der Hucht (2001), 
(3) GOS Catalog V2.2, Sota et al. (2008), (4) Massey (1981), 
(5) Landolt-R$\ddot{o}$rmstein (1996), (6) Wegner 1994, (7) SIMBAD database, 
(*) Spectral classsification B0III-B2V was adopted, (**) spectral class V 
and III vas adopted.}
\end{center}
\end{table*}  

New H$\alpha$ and [O III] images taken by  Miller \& Chu (1993) revealed 
an arc of 4\arcmin\ in size to the south of the WR star (at 
$[l,b]$ = [102\gra 46\arcmin, --0\gra 43\arcmin]). They concluded that 
the filament probably indicates the position of an ionization front and 
considered it to be a ring nebula related to the star. Based on optical
data, Esteban \& Rosado (1995) found velocities in the range --41 to --59 
\kms\ for the arc and derived an electron density of 290$\pm$100 cm$^{-3}$ 
from line ratios. The [OIII]/H$\alpha$ ratio is consistent with values derived 
for ring nebulae, suggesting that WR\,153ab is the main source of ionization 
(Esteban \& Rosado 1995). These authors classify the arc as $R_s$-type based on the morphology of this feature and on the absence of a clear evidence of expanding motions in the line profiles. This last point suggests that the massive progenitor of the WR star contributed in the shaping of the nebula (see Chu et al. 1981).

Distance estimates for the WR star based on different methods range from
2.75 to 5.0 kpc. Massey (1981) derived a spectrophotometric distance of 
5.0 kpc, while van der Hucht (2001) estimated 2.75 kpc based on its 
association to Cep\,OB1, with a distance uncertainty of 40\%. 
Foster \& Routledge (2003), using a method based on \hi\ column 
densities, estimated 3.2$\pm$0.5 kpc. A distance of 3.68 kpc was adopted by 
Churchwell \& Wamsley (1973), based on the WN6+BOI spectral type of the WR 
star, known at that time. Georgelin \& Georgelin (1976) derived a 
distance of 3.6 kpc, while Fich, Blitz \& Stark (1989) obtained 4.2$\pm$1.5 
kpc.

Adopting a mean LSR velocity of --48 \kms\ for the ionized gas, and taking
into account the presence of non-circular motions in this section of the
Galaxy (see Fig. 2a by Brand \& Blitz 1993), a kinematical distance 
$d_k$= 3.5$\pm$1.0 kpc can be predicted, in close agreement with 
distances derived optically for the WR star. We adopt this distance in 
the present study.

With the aim of investigating  in some detail the nature of Sh2-132, we 
analyzed the distribution of the different gas components in its environs.
To accomplish this goal we performed a multi-frequency study using radio 
continuum data at several frequencies, \hi\ 21cm line data, $^{12}$CO(1-0) 
line data, and IR observations at different wavelengths. These data revealed
that photodissociation regions are present at the interface between the
ionized and molecular gas.

\section{Database}

Neutral hydrogen 21cm line data from the Canadian Galactic Plane Survey 
(CGPS, Taylor et al. 2003) was used to analyze the \hi\ gas distribution, 
while $^{12}$CO(1-0) data at 115 GHz from the Five College Radio Astronomy 
Observatory, allowed us to study the molecular distribution. 
Intensities are shown as main beam brightness temperatures $T_{mb}$. The \hi\ data were smoothed to $1\farcm 4\times1\farcm 4$ to facilitate the identification of the \hi\ features. The main observational 
parameters of these data bases are detailed in Table~\ref{cubos}.

Radio continuum observations at 408 and 1420 MHz were also extracted from 
the CGPS. The image at 408 MHz was obtained with a synthesized beam and an
rms noise of 3\farcm 4$\times$2\farcm 8 and $\sim$1.1 K, respectively; 
the corresponding values for 1420 MHz are 1\farcm  0$\times$0\farcm 81, and 
0.063 K.

Mid-infrared (MIR) images were extracted from the MSX Galactic Plane Survey 
(Price et al. 2001) at 8.3$\mu$m  (A-band), 12.1$\mu$m  (C-band), 14.7$\mu$m  
(D-band), and 21.3$\mu$m (E-band), with an angular resolution of 18\farcs 4. MIR and far-infrared (FIR) images at 12, 25, 60, and 100 $\mu$m from the IRAS satelite (HIRES data) are also included. These images have angular resolutions in the range 0\farcm 5 to 2\farcm 0.

\begin{table}
 \centering
  \begin{minipage}{140mm}  
\caption{Observational parameters of the \hi\ and $^{12}$CO data}
\label{cubos}
 \begin{tabular}{lcc}
\hline
    &  \hi\ &  $^{12}$CO(1-0) \\
\hline
Original FWHM  & 1\farcm 18$\times$0\farcm 98  &   $\sim$1\arcmin \\
Final FWHM  & 1\farcm 4$\times$1\farcm 4   \\
Vel. range (LSR) (\kms)  & --165,+57 & --165, +57 \\
Vel. resolution (\kms)   & 1.32 &  0.824 \\
RMS noise $T_b$ (K) & 1.7 & 0.15  \\
\hline
\end{tabular}
\end{minipage}
\end{table}

\section{Results}

\subsection{Radio continuum emission}\label{continuo}

\begin{figure}
\includegraphics[width=84mm]{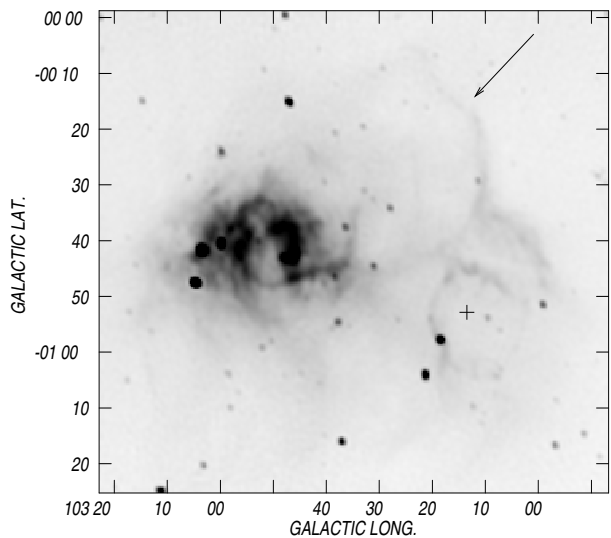}
\caption{Full resolution CGPS image at 1420 MHz emphasizing faint emission 
to the north of WR~152 (indicated by {\bf a cross symbol} in the lower right part 
of the figure). The arrow points to a continuum feature corresponding to 
the large HI structure.}
  \label{figlarge}
\end{figure}

\begin{figure}
    \includegraphics[width=84mm]{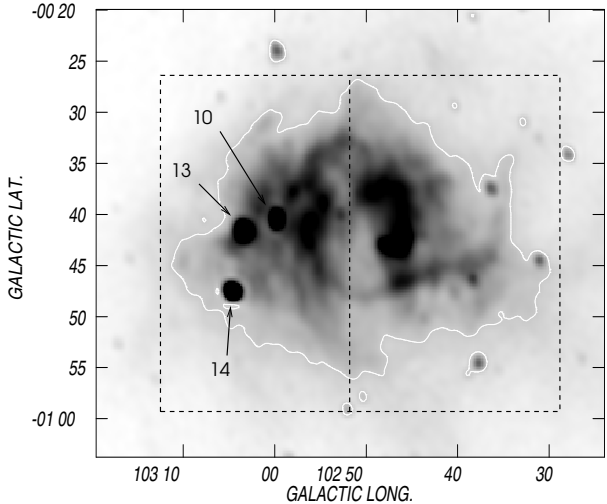}
\caption{Full resolution 1420 MHz CGPS image showing the eastern and 
western regions of Sh2-132 used for TT-plot analyses. The radio sources in direction to Sh2-132 are indicated by 10, 13 and 14. The white contour 
line marks the 9 K contour of the convolved 1420 MHz image used for the 
analysis.  See text for details.}
  \label{figwherett}
\end{figure}

Figures~\ref{figlarge} and \ref{figwherett} show the CGPS radio continuum 
emission at 1420 MHz. The grayscale in Fig. 2 has 
been chosen to emphasize the large-scale region of faint emissivity to the 
north of WR\,152 (Cappa et al. 2010, hence the ``burnt-out'' appearance 
of the region around WR~153ab). 
The emission distribution at 1420 MHz resembles an upside-down ``U''. 
This semicircular structure consists of two areas of strong emission 
coincident with Shells A and B detected by Harten et al. (1978),
centered at $(l,b)$ $\simeq$ (102\gra 34\arcmin 42\farcs 4,
--00\gra 46\arcmin 9\farcs 8) and at $(l,b)$ $\simeq$ 
(102\gra 26\arcmin 35\farcs 8,--00\gra 43\arcmin 10\farcs 2), respectively.
WR\,153ab, indicated by the larger cross in Fig. 1, 
appears projected onto the central region of Shell B.  
Thus, the bright radio emission at $l\geq$ 102\fdg 5 is in excellent 
agreement with the optical emission (see Fig.~\ref{opt1-sh2-132})  
and with 
the image at 610 MHz by Harten et al. (1978). Both Shells A and B can be 
identified at 1420 MHz. Note in particular the small arc of emission at 
$(l,b)$ = (102\fdg 77, --0\fdg 73) (best identified in Fig. 3), which is 
the radio counterpart of the 
4\arcmin-arc detected at optical wavelengths by Miller \& Chu (1993).

The faint structure centered at $(l,b)$ = (102\fdg 2, --0\fdg 83) is related 
to WR\,152 and has been analyzed in a previous paper (Cappa et al. 2010).  
Of particular interest is the curved  filament indicated by an arrow in Fig. 
2. This filament 
can be followed from $b \approx$ --0\fdg 75 to nearly $b \approx$ --0\fdg 05.
The figure also shows a faint barely visible northeastern section.
This structure is likely the radio continuum counterpart of the 
$\sim$60\arcmin\ diameter \hi\ structure discussed in Cappa et al. (2010).

Three bright point sources, best seen on Figure~\ref{figwherett}, are projected onto the nebula near Shell A. Their 
coordinates, flux densities and identifications are included in Table 3.
The spectral index estimates indicate that sources 10 and 14 are 
non-thermal in
nature and probably extragalactic, while source 13 is thermal. The value of
the spectral index indicates that the radio source is optically thick at
least at 408 MHz.

The CGPS image at 408 MHz does not show as many details as the one
at 1420 MHz, as it is of much lower resolution.  However it can be
used to study the spectral index distribution of the radio continuum
emission around WR~153ab. As a first step, we removed the point sources which 
are present over the region of interest and convolved the
1420 and 408 MHz images to the same spatial resolution of 3\farcm 4.
We then constructed a TT-plot, in which the brightness temperature $T_b$ 
at one frequency is plotted point-by-point against the brightness 
temperature at the other frequency. The brightness temperature spectral 
index $\beta$, where $T_b \propto \nu^{\beta}$, is directly related to the 
slope of the straight line fitted by regression,
$\beta = -\,\log(\rm slope)/\log(1420/408)$. The usual flux density spectral
index $\alpha$ ($S_{\nu} \propto \nu^{\alpha}$) is simply 
$\alpha = \beta + 2$. The TT-plot method is immune from errors in the 
zero-level of the individual images (which the simpler technique of dividing 
one map by the other, even after background subtraction, is not).

Three different TT-plots were constructed, one for the entire Sh2-132 region 
and one for the eastern and western halves, which correspond to the
two shells identified by Harten et al.~(1978). 
Figure~\ref{figwherett}, which is the full resolution image at 1420 MHz,
shows these different regions,  the dashed lines delimiting the eastern and 
western parts. The three bright small diameter sources coincide with 
sources 10, 13 and 14 in Table 2 of Harten et al.~(1978) and had been 
subtracted from the original image before convolution to a circular beam 
of 3\farcm 4.

\begin{figure}
    \includegraphics[width=70mm]{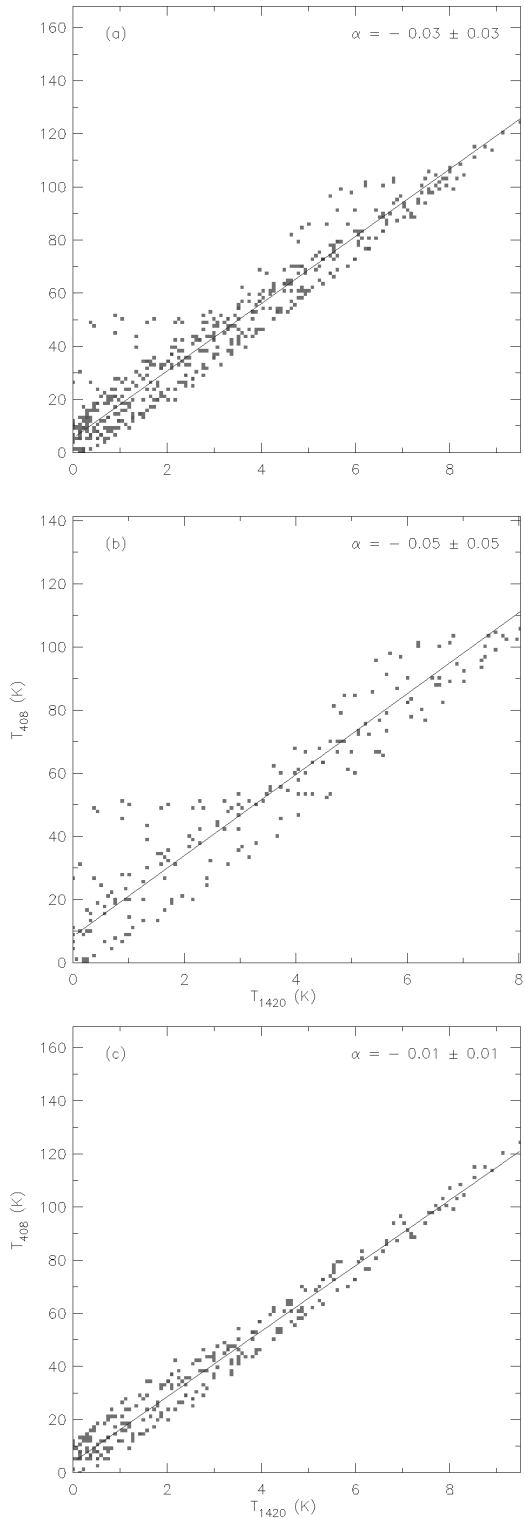}
\caption{TT-plots corresponding to (a) the whole structure, (b) the eastern 
half and (c) the western half.  For all plots, a background of 9 K and 
80 K has been subtracted from the convolved images at 1420 and 408 MHz, 
respectively.}
  \label{figtt}
\end{figure}

The results of the TT-plot analysis are presented in Figure\ref{figtt}. 
In each case, the analysis was limited to regions brighter than 9 K at 1420 
MHz and 80 K at 408 MHz. Although a significant amount of scatter is present 
on the diagram, especially for the eastern sector, there is no evidence 
for the presence of significant variations of the spectral index and a 
further study of spatial variations is not warranted.  The best-fit 
spectral index is $\alpha$ = --0.03$\pm$0.03, corresponding to thermal 
emission. 

In order to obtain some radio continuum parameters, we first derive
the total number $N_u$ of UV ionizing photons per second required to produce
the observed free-free emission. This can be written as (e.g. Chaisson 1976):
\begin{equation}
N_u  = 0.76 \times 10^{47}\,\,T_4^{-0.45}\,\nu_{\rm GHz}
           ^{0.1}\,S_{\nu}\,\,d_{\rm kpc}^{2},
\end{equation}
where $S_{\nu}$ is the flux density in Jy, $T_4$ is the electron temperature
in units of $10^4$ K and $\nu_{\rm GHz}$ the frequency in GHz. We use
the CGPS image at 1420 MHz.  The derived
flux density is significantly dependent on how the source dimension is
defined. The surrounding background is somewhat variable, nevertheless
it is reasonable to assume that the source emission corresponds to
brightness temperatures above 6.5 to 7 K.  Using an average background of
7 K, we obtain a flux density at 1420 MHz of $23 \pm 5$ Jy, the error being
determined from a second estimate using a background of 6.5 K. From
the above equation for $N_u$ and taking $T_4 = 0.8$ (Quireza et al. 2006), 
we obtain $N_u = 2.2 \times 10^{49} \,\rm s^{-1}$. Considering two O6I-type 
stars and two WN6 as components of the multiple system, and the spectral 
type from BD+55\gra 2722, LS\,III+55\gra 45, and LS+55\gra 40 (see Table 1, 
related to the \hii\ region, we obtain the UV ionizing photons per second 
emited by the stars, $Q_0$=(16.6$\pm$3.0)$\times$10$^{49}$ s$^{-1}$, for 
solar metalicity from Smith et al. (2002).

The comparison between the number of UV photons emitted by the stars $Q_0$
and the number of UV photons used to ionize the gas $N_u$ indicates that 
the massive stars in the nebula are responsible for the ionization. 
A large fraction of the UV photons are also used to heat the associated
interstellar dust. This result differs from the one by Harten et al.
(1978) since their estimate of  $Q_0$ was too low. 

\begin{table*}
\caption{Small diameter radio sources towards Sh2-132.}
\label{tabla2}
\begin{tabular}{ccccccccccl}
\hline\hline
\# &$l$  & S$_{1420}$ & S$_{408}$ & $\alpha$ & S$_{178}$ & 
S$_{365}$ & S$_{610}$ & S$_{1420}$ & S$_{4850}$ & Identification \\
 & $b$ &  mJy & mJy &  & mJy  &  mJy & mJy  & mJy   \\
 (1) &   & (2)  & (2)  & (2) & (3) & (4) & (5) & (6) & (5) \\
\hline\hline
10 & 102\gra 59\arcmin 50\farcs 90 & 
195$\pm$10 & 550  & --0.8  &  & 819$\pm$ 87 & 330  & &  $<$40 &  7C 2218+558$^{(7)}$  \\
 & --0\gra 40\arcmin 23\farcs 35  & \\
13 & 103\gra 3\arcmin 30\farcs 24  & 
365$\pm$15 & 280  & +0.2    &  &  & 160  &  222.6$\pm$7.5 & $<$40 & NVSS J222034+561438  \\
 & --0\gra 41\arcmin 28\farcs 14 \\
14 & 103\gra 4\arcmin 40\farcs 04 & 450$\pm$20 & 1200  & --0.8    & 3100$\pm$500 & 1738$\pm$62 & 810  &  & 60 &  TXS 2219+559  \\
 & --0\gra 47\arcmin 27\farcs 04 \\
\hline
\end{tabular}
References: (1) Number from the list of Harten et al. (1978); (2) this paper, 
derived using CGPS data; (3) Gower et al. (1967);
(4) Douglas et al. (1996); (5) Harten et al. (1978); (6) Condon et al. (1998); 
(7) Waldram et al. (1996).
\end{table*}

We can now use the models of Mezger \& Henderson (1967) to infer the 
properties of the extended ionized gas, namely its mass
$M_{\rm HII}$ and the rms electron number density $n_e$.  
Letting $\theta_{Ga}$ be the observed half-power width of the
extended emission (in minutes of arc, after correction for the beam),
we obtain:
\begin{equation}
n_e/{\rm cm^{-3}} = 6.35 \times 10^2 \,u_1\,a^{-1/2}\,T_4^{0.175}\,\nu_{\rm GHz}
           ^{0.05}\,S_{\nu}^{0.5}\,\theta_{\rm Ga}^{-1.5}\,d_{\rm kpc}
           ^{-0.5}
\end{equation}
\begin{equation}
M_{\rm HII}/M_{\odot}=0.386\,u_2\,a^{-1/2}\,T_4^{0.175}\,\nu_{\rm GHz}
           ^{0.05}\,S_{\nu}^{0.5}\,\theta_{\rm Ga}^{1.5}\,d_{\rm kpc}
           ^{2.5}
\end{equation}
where $a(\nu,T_e)$ is the Gaunt factor (about unity for our purposes),
$d_{\rm kpc}$ is the distance in kpc
and the parameters $u_1$ and $u_2$, both of order unity, depend on the
assumed model for the source. These parameters are given by Mezger \&
Henderson (1967) for three different idealized models. From Figure\ref{figwherett}, it is apparent that the nebulosity can be roughly
broken up into two different regions separated by a nearly vertical line
running at a longitude of $\sim$102\fdg 87.  
These two regions correspond to shells A and B from Harten et al.~(1978).
The flux densities of the two sub-regions are $11 \pm 2$ Jy and $12 \pm 3$ 
Jy, respectively. For simplicity, we model each half-region as a sphere 
of constant density and of angular size 12\arcmin. From Mezger \& 
Henderson (1967), we find $u_1 = 0.78$, $u_2 =1.29$.
Assuming $T_4 =0.8$, from which we determine $a^{-1/2} =  1.04$,
we finally have an rms $n_e = 20\,\rm cm^{-3}$ and $M_{\rm HII} = 
1500 M_{\odot}$ for each region, where, given the error in 
estimating fluxes, we have simply 
taken $S_{\nu} = 10 \,\rm Jy$ for each half.
The derived rms electron density is compatible with estimates 
by Harten et al. (1978) for Shells A and B.

\begin{figure}
    \centering
    \includegraphics[width=84mm]{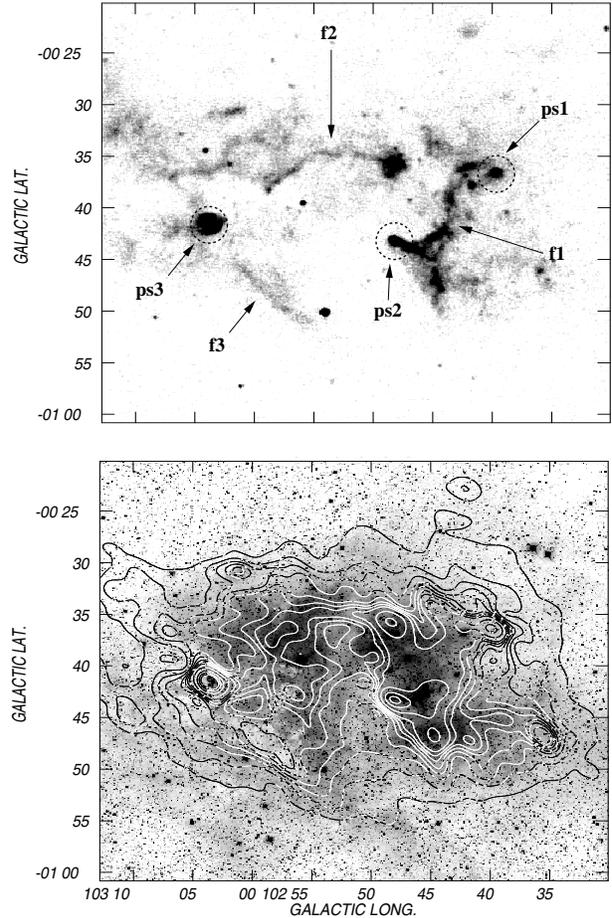}
       \caption{{\it Upper pannel} Sh2-132 at 8.3 $\mu$m (MSX band A). 
The grayscale goes from 8.0$\times$10$^{-7}$ to 3.5$\times$10$^{-6}$ W 
m$^{-2}$ ster$^{-1}$. 
Extended filaments and point-like sources are indicated (see text).
{\it Bottom pannel:} Overlay  of the DSS-R image  and 
the image at 60$\mu$m. IR contours  are 50 to 190 MJy ster$^{-1}$ 
in steps of 20 MJy ster$^{-1}$, and from 300 to 900 MJy ster$^{-1}$ 
in steps of 200 MJy ster$^{-1}$.}
 \label{fig:imagenesIR}
 \end{figure}

\subsection{IRAS and MSX emission}

The IRAS (60$\mu$m) and MSX band A (8.3$\mu$m) emissions are shown in 
Fig.\ref{fig:imagenesIR}. 
Although we analyzed the four IRAS bands, we
include the 60 $\mu$m-image only since the extended emission 
is clearly seen at this wavelength. Emission at 8.3$\mu$m (upper panel) in diffuse nebulae has a contribution from both PAH and dust continuum. As pointed out by Povich et al. (2007), in very energetic regions like M\,17, dust continuum can be very strong. However, for the case of Sh2-132, the assumption that MSX band A is dominated by PAH emission seems to be reasonable. Consequently, emisison in this band indicate the presence of photodissociation regions, while the emission
in the far IR originates in large dust grains.

 The image at 8.3 $\mu$m (upper panel) reveals the existence of a ring-like 
feature centered at $(l,b)$ = (102\gra 53\arcmin, --00\gra40\arcmin) of 
about $\sim$15\arcmin\ in size. This feature can be also identified at
60 $\mu$m (lower panel). The comparison of the optical emission with the 
emission at 8.3 $\mu$m shows that the mid IR emission encircles most of
the nebula, following the outer rim of the ionized gas. 
This fact is more noticeable for Shell B and is 
particularly striking near $(l,b)$ = (102\gra 47\arcmin,--00\gra 43\arcmin), 
where the 4\arcmin\  ring nebula is present. 
The bottom panel shows that the emission distribution at 60 $\mu$m 
resembles that in the optical.

In order to estimate some dust parameters, we have divided this ring-like
structure in three main filaments, named f1, f2, and f3, which are indicated 
in the upper panel of Fig. 5. Note that f3 borders the region with high 
extinction
seen in the optical image near $(l,b)$ = (103\gra 0\arcmin,--00\gra 47\arcmin).

A number of interesting point-like sources located onto the filaments can 
also be identified in the MSX and IRAS images. They are indicated 
in the figure as ps1, ps2, and ps3. Infrared source ps1 at $(l,b)$ = 
(102\gra 39\arcmin 45\arcsec,--00\gra 36\arcmin 24\arcsec)
is the MSX source G102.6627-00.6067, coincident with IRAS\,22160+5551; ps2
at $(l,b)$ = (102\gra 48\arcmin 18\arcsec,--00\gra 43\arcmin 6\arcsec) 
corresponds to  G102.8051-00.7184 and IRAS\,22172+5549; while ps3 at $(l,b)$ =
(103\gra 03\arcmin44\arcsec,--00\gra 41\arcmin 27\arcsec) is 
G103.0624-00.6911 and IRAS\,22187+5559.

After background substraction, the IR flux densities of the 
filaments in the four IRAS bands and in the MSX band A were 
calculated. These values are listed in Table~\ref{table:flujos}. 
The quoted errors reflect the uncertainties in the determination of the 
background. Emission in bands C, D, and E is very weak in comparison with
the strong background emission leading to large uncertainties in flux 
density estimates. Values for these bands are not included.
The three point sources, whose flux densities in the MSX and IRAS 
bands are included in the table, can be classified as YSO candidates following 
criteria by Junkes et al. (1992) and Lumsden et al. (2002) for IRAS and
MSX point sources, respectively. In the case of the IRAS criteria, Junkes et al. (1992) consider (i) $S_{100}>$20 Jy, (ii) 1.2$<S_{100}/S_{60}<$6, (iii) $S_{60}/S_{25}\geq$1, and $Q_{60}$+$Q_{100}\geq$4, where $S_i$ is the flux density in the bands centered at 25, 60 and 100 $\mu$m, and $Q_i$ is the quality of the fluxes at these bands. Lumsden et al.'s (2002) criteria are (i) $S_{21}/S_8>2$ and (ii) $S_{14}/S_{12}>1$, where $S_i$ is the flux density centered at 8.3, 12.1, 14.7, and 21.3 $\mu$m.

\begin{figure}
\centering
\includegraphics[width=84mm]{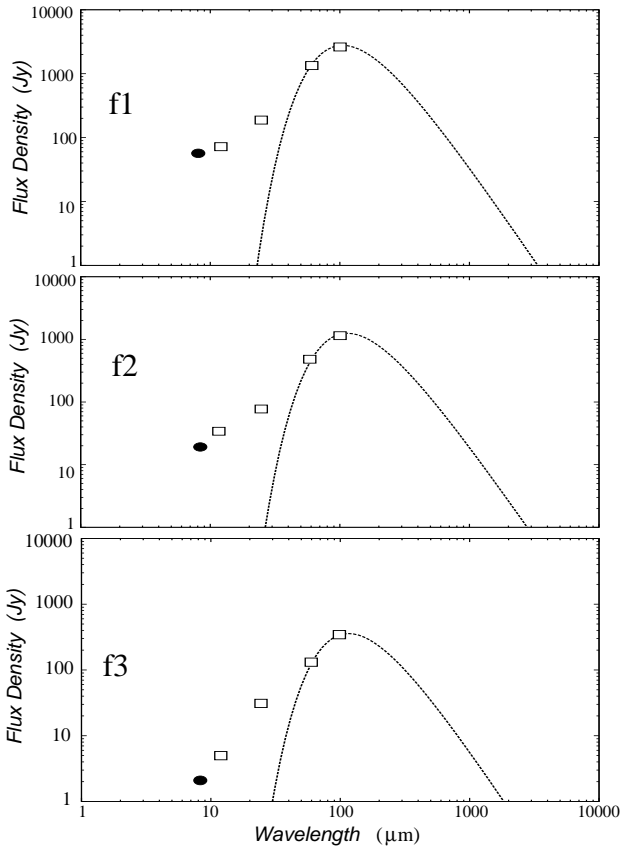}
\caption{IR spectral energy distribution for the IR  filaments.  
The  lines in the diagram corresponding to $f1$, $f2$, and $f3$ 
show the emission of a blackbody with a {\bf $T$ = 32, 29 and 32 K,
 respectively,}  and a {\bf $\nu^{1.5}$}-emissivity law. The boxes denote flux 
densities in the four IRAS bands, and the circle, that corresponding to the 
MSX band A.}
\label{fig:sed}
\end{figure}

It has been demonstrated (Scoville \& Kwan 1976) that the infrared spectrum 
of radiatively excited dust grains can be fitted by a modified Planck 
function $B(T_d)\nu^m$, were $\nu^{m}$ is the grain emissivity and $m$ 
increases from 1 to 2 depending on the wavelength and composition of dust 
grains (Schwartz 1982). Adopting $m$ = 1.5, according with the molecular environment (Whittet, 1992), and using  the measured flux 
densities at 60 and 100 $\mu$m, we derived the dust color temperature 
T$_d$ summarized in Table~\ref{table:flujos}. IR luminosities for the point sources are also included. Infrared 
filaments are excited by the intense stellar radiation field of WR\,153ab 
and the other massive stars in the region. 
The slightly higher dust color temperature derived for f1 can be
explained bearing in mind that this filament is closer to the 
exciting stars than f2 and f3.
In Fig.~\ref{fig:sed}, 
the MIR and FIR spectral energy distribution 
corresponding to the filaments are shown. Note that the measured fluxes 
at shorter wavelengths exceed the values of our model, possibly as the 
result of the emission of nebular lines which may contribute to the 
infrared flux densities.

\begin{figure}
\includegraphics[angle=0,width=84mm]{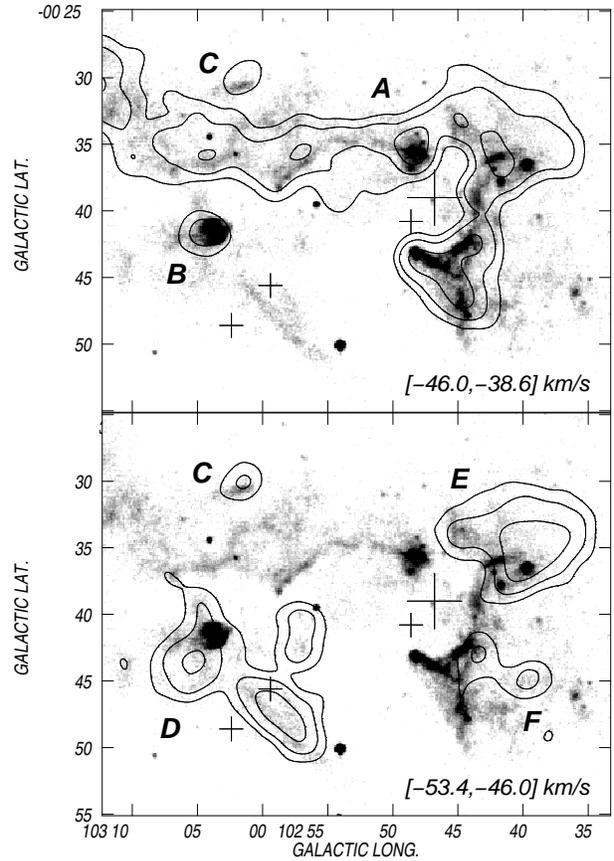}
\caption{Overlay of the averaged $^{12}$CO emission distribution for the 
molecular clouds between --46.0 to --38.6 \kms\ (upper panel) and --53.4 to 
--46.0 \kms\ (lower panel) and the MSX band A emission (grayscale). 
Contours correspond to 3, 7, and 15$\sigma$. The grayscale goes from
0.1$\times$10$^{-7}$ to 1$\times$10$^{-7}$ Watts m$^{-2}$ ster$^{-1}$. 
The crosses indicate the positions of the massive stars.}
\label{co-msx}
\end{figure}
\begin{figure}
\centering
\includegraphics[angle=0,width=84mm]{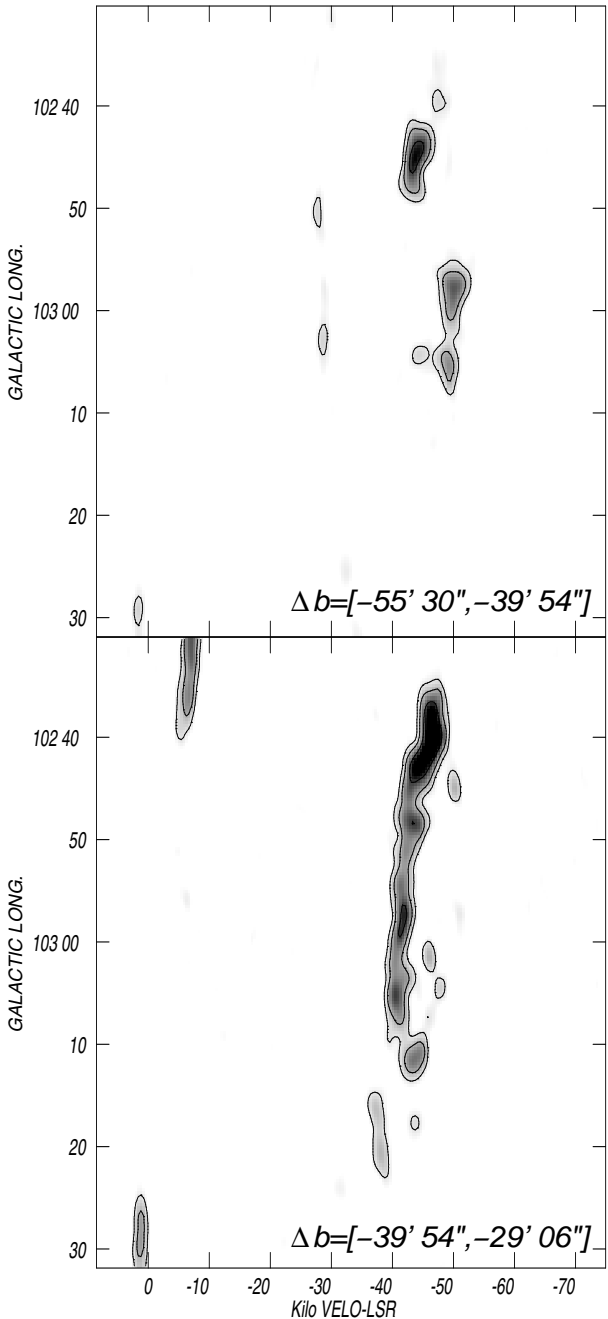}
\caption{{\it Upper panel:} Averaged $^{12}$CO velocity distribution 
for  $\Delta b$ = (--55$\arcmin$30$\arcsec$,--39$\arcmin$54$\arcsec$). 
The grayscale goes from 0.15 to 3 K. Contours are 0.45, 0.75 and 1.05 K.
{\it Bottom panel:} Averaged $^{12}$CO velocity distribution between 
$\Delta b$ = (--39$\arcmin$54$\arcsec$,--29$\arcmin$06$\arcsec$). 
Contours and grayscale are the same as in the upper panel.}
\label{co-v-l}
\end{figure}

\begin{table*}
\begin{center}
\begin{minipage}{17cm}
\caption{IR filaments and point sources.}
\label{table:flujos}
\begin{tabular}{lcccccccccc}
\hline
 Source & \multicolumn{8}{c}{Fluxes [Jy]}  &  $T_d$ (K)& $L_{IRAS} (L_{\odot})$ \\
        & 8.3 $\mu$m & 12.1 $\mu$m & 14.7 $\mu$m & 21.3 $\mu$m & 12 $\mu$m  & 25 $\mu$m & 60 $\mu$m    & 100 $\mu$m  \\
\hline\hline
f1      & 46$\pm$1   &  ---        & ---         & ---         & 72$\pm$4   & 192$\pm$24& 1370$\pm$130 & 2765$\pm$145 & 32$\pm$7  & 43800   \\ 
f2      & 16.0$\pm$3.5  &  ---        & ---         & ---         & 35$\pm$6   & 77$\pm$11 & 484$\pm$51   & 1201$\pm$160 & 29$\pm$6  & 16900    \\     
f3      & 1.8$\pm$0.3& ---         & ---         & ---         & 4.9$\pm$0.5 & 32$\pm$11& 133$\pm$40   & 343$\pm$37   & 32$\pm$6  &5000     \\         
ps1     & 1.45       & 1.85        & 0.97        &  2.14       &  3.03       & 4.28     & 74.1         & 274          & 28$\pm$5 &2700      \\      
ps2     & 2.22       & 3.93        & 3.78        &  7.02       &  7.8        &  16.1    & 106          & 318          & 30$\pm$5 &  3900    \\
ps3     & 1.0        & 1.6         & 1.40        &  4.26       &  7.02       &  17.2    & 243          & 453          & 36$\pm$8 &  6800     \\
\hline
\end{tabular}
\end{minipage}
\end{center}
\end{table*}

\subsection{Molecular gas distribution}\label{CO}

The analysis of the $^{12}$CO(1-0) emission distribution leads to the 
detection of two structures with velocities in the ranges from --53.4 to
46.0 \kms, and from --46.0 to --38.6 \kms. Fig.~\ref{co-msx} shows the 
averaged CO emission (contours) for the two molecular structures superposed 
onto the emission at 8.3 $\mu$m. 

The CO feature present between --46.0 and --38.6 \kms\ (upper panel) exhibits 
a ring-like appearance between $l\sim$ 102\gra 35\arcmin\ and  102\gra 
50\arcmin, while for $l>$ 102\gra 45\arcmin, the emission is concentrated 
along $b\simeq$ --0\gra 35\arcmin. The whole structure will be referred as 
cloud A. 
Two smaller molecular clouds are detected  at $(l,b)\sim$ 
(103\gra 2\arcmin,--0\gra 30\arcmin) (indicated in the figure as cloud C) 
and at $(l,b)\sim$ (103\gra 4\arcmin,--0\gra 42\arcmin) (cloud B). 
The emission at 8.3 $\mu$m correlates very well with the molecular emission.
In particular, the brighter emission regions in the mid infrared coincide
with clumps in cloud A [for example at $(l,b)\sim$ 
(102\gra 43\arcmin,--0\gra 35\arcmin), (102\gra 44\arcmin,--0\gra 43\arcmin), 
(102\gra 47\arcmin,--0\gra 44\arcmin), and 
(103\gra 5\arcmin,--0\gra 42\arcmin)]. Cloud B coincides with the strong 
radio source NVSS J222034+561438, which is an YSO candidate. The comparison 
of the ionized and molecular gas distributions shows that the molecular 
emission encircles the brightest optical emission regions analyzed by 
S\'anchez-Monge et al. (2008).

The lower panel displays the molecular emission between --53.4 to --46.0  
\kms. The CO emission is clumpy, being concentrated in clouds D, E and F. 
Cloud C is also detected in this velocity range. Cloud E overlaps a section of
cloud A and follows the ring-like feature identified at 8.3 $\mu$m. 
Cloud F coincides with part of the IR filament at $(l,b)\sim$ 
(102\gra 44\arcmin,--0\gra 42\arcmin). The section of Cloud D running from
$(l,b)$ = (102\gra 56\arcmin,--00\gra 51\arcmin) to $(l,b)$ = 
(103\gra 02\arcmin,--00\gra 44\arcmin) coincides with f3. Cloud D also
displays a clump at $(l,b)\sim$ (103\gra 05\arcmin,--0\gra 44\arcmin),
which coincides with a region of high optical extincion and 
is probably connected with cloud B. 

Table~\ref{tabla_co} gives the physical parameters of each molecular cloud.
The masses of the different clouds were estimated by integrating the CO 
line intensity as $W(CO) = \int T_{mb}dv$ = $T_{mb}\Delta v$, 
where $T_{mb}$ is the average main 
beam brightness temperature of the molecular cloud over the velocity 
interval in which the cloud is observed. To estimate the $H_2$ column 
density $N(H_2)$, the relation $X = N(H_2)/W(CO)$ = 
(2.3$\pm$1.2)$\times$10$^{20}$ 
mol cm$^{-2}$ K$^{-1}$ km$^{-1}$ s (Grenier \& Lebrun 1990) 
was used. The molecular mass was derived from $M$(\Msun) = 
4.2$\times$10$^{20}N(H_2)d^2 \Omega$, where $d$ is the distance in pc and 
$\Omega$ is the solid angle in steradians. Errors in $T_{mb}$ and
in $N(H_2)$ arise in background uncertainties, while errors in masses 
come from background and distance uncertainties. The adopted
distance is 3.5$\pm$1.0 kpc.

To determine the volume density, we assumed a spherical geometry for all
the clouds, except for Cloud A, for which  an elongated geometry was 
adopted.

\begin{table*}
\begin{center}
\begin{minipage}{17cm}
\caption{Physical parameters of the molecular clouds.}
\label{tabla_co}
 \begin{tabular}{lcccccc}
\hline
          &    \multicolumn{6}{c}{Clouds} \\
         & A & B & C & D & E & F  \\
\hline
Average  $T_{mb}$ (K) &  1.30$\pm$0.30& 0.90$\pm$0.15&  0.60$\pm$0.15 & 1.20$\pm$0.30& 1.60$\pm$0.15&0.70$\pm$0.15\\
Velocity range (\kms )  & --38.6,--46.0& --38.6,--46.0 &--38.6,--53.4 &--46.0,--53.4 &--46.0,--53.4 &--46.0,--53.4\\
$N(H_2)$ column density (10$^{20}$\cmdos )  &6.50$\pm$1.60 & 3.30$\pm$0.60& 4.20$\pm$1.05 & 5.90$\pm$2.40 & 9.0$\pm$0.90 &1.95$\pm$0.20\\
$H_2$ mass (10$^3$ \Msun ) & 4.25$\pm$1.90 & 0.80$\pm$0.35&  0.04$\pm$0.02 & 1.30$\pm$0.60&1.55$\pm$0.70 & 0.95$\pm$0.45\\
Volume density (\cm ) & 37$\pm$22 & 2150$\pm$1290 & 110$\pm$65 & 95$\pm$55 & 115$\pm$70 & 320$\pm$190 \\
$N(HI)$ column density (10$^{21}$\cmdos )   & 3.7$\pm$0.9 & 3.5$\pm$0.5 & 3.7$\pm$0.9 & 3.3$\pm$0.8  &  3.4$\pm$0.3  & 3.3$\pm$0.6 \\
Visual absorption $A_v$ (mag) & 2.5$\pm$0.8 & 2.2$\pm$0.5 & 2.8$\pm$0.7  & 2.4$\pm$0.7 & 2.8$\pm$0.6 &2.7$\pm$0.5 \\
\hline
\end{tabular}
\end{minipage}
\end{center}
\end{table*}

Figure~\ref{co-v-l} shows two velocity-position maps of the CO emission. 
We have plotted Galactic longitude in the y-axis since the molecular 
morphology is mostly elongated in this direction. The map in the upper 
panel is the result of integrating within $\Delta b$ = 
\hbox{(--0\gra 55\arcmin 30\arcsec,--0\gra 39\arcmin 54\arcsec)}, while the map 
in the bottom panel was integrated within $\Delta b$ = 
(--0\gra 39\arcmin 54\arcsec,--0\gra 29\arcmin 06\arcsec). 
The upper panel shows two clouds at $l \sim$ 102\gra 45\arcmin\ and 
$l \sim$ 103\gra 00\arcmin, having velocities in the range $\Delta v \sim$ 
(--40,--45) \kms\ and (--45,--50) \kms, respectively (two small cloudlets
are also detected at $v\simeq$ --30 \kms\ at $l \sim$102 \gra 50\arcmin\  
and $l \sim$103 \gra 03\arcmin). Clouds F and D are clearly identified. 
The bottom panel shows cloud A extending from --102\gra 35\arcmin\ to 
103\gra 13\arcmin. The velocity range goes from --40 \kms\ at $l \sim$
103\gra 10\arcmin\ to --50 \kms\ at $l \sim$ 102\gra 35\arcmin. 
The image shows some small cloudlets at $l \sim$ 103\gra 23\arcmin\ 
at $v\simeq$ --38 \kms,   $l<$ 102\gra 40\arcmin\ with velocities in
the range $\sim$ --8 to --5 \kms, and at $l >$ 103\gra 25\arcmin\ at 
$v\simeq$ 0 \kms. Molecular gas with velocities higher than --10 \kms\
is unconnected to Sh2-132. 

The agreement in velocity between the ionized and molecular gas and the 
fact that the molecular emission borders the brightest ionized regions 
indicate that molecular gas with velocities between --38 and --54 \kms\
is associated with Sh2-132 and the massive stars in the region. 

We adopt --45 \kms\ as the systemic velocity of the molecular gas.
Taking into account the non-circular motions in this section of the Galaxy
shown by Brand \& Blitz (1993), a kinematical distance $d_k$ = 3.0$\pm$1.0 
kpc can be estimated, compatible with the distance adopted in 
Sect.\ref{intro}.

\subsection{The distribution of the \hi\ emission }\label{HI}

\begin{figure}
\centering
\includegraphics[angle=0,width=84mm]{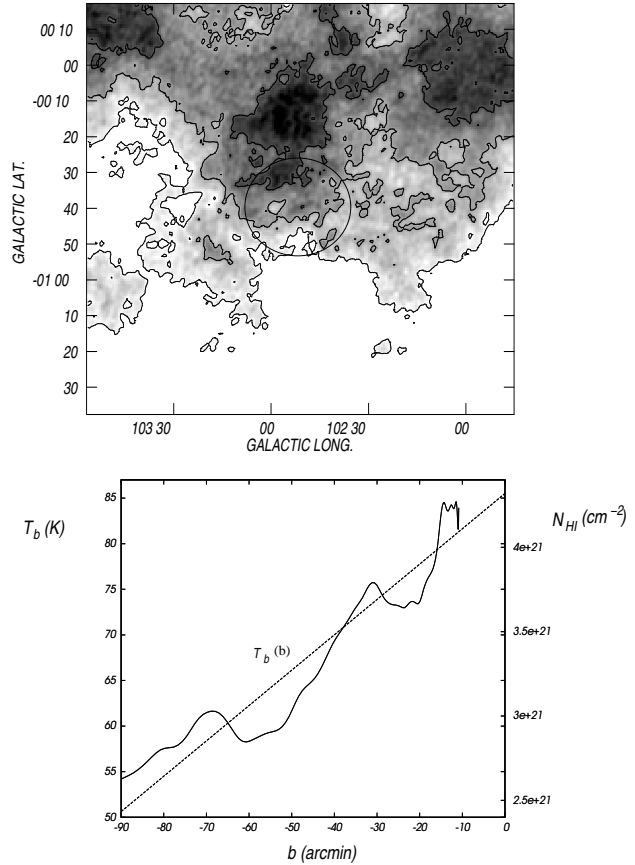}
\caption{{\it Upper panel:} Average \hi\ emission distribution between 
--31.1 and --58.4 \kms. 
The grayscale goes from 50 to 93 K, while contours correspond to 60, 70, 80,
and 90 K. The white circle has the dimensions of the \hii\ region Sh2-132.
{\it Bottom panel: } Diagram showing $T_b$ vs. $b$ from $b$= --90$\arcmin$ 
to 0$\arcmin$ for $l$ = 102\gra 50\arcmin\ from the image above. The dashed 
line shows the best fit to the data.}
\label{hi-corte}
\end{figure}

The upper panel of Fig.\ref{hi-corte} 
displays the \hi\ emission 
distribution in the velocity range where the CO emission is present, i.e. 
between --31.1 and --58.4 \kms. The white circle has the dimensions of the 
\hii\ region.  Neither a region lacking \hi\ gas coincident in position with 
Sh2-132 as a whole nor an \hi\ shell surrounding it are evident in the 
image, although the small cavity near $(l,b) \simeq$ 
(102\gra 58\arcmin,--0\gra 40\arcmin) is probably linked to shell A.

However, a  gradient in the \hi\ emission distribution is evident for
$b <$ --0\gra 10\arcmin. The lower panel displays the emission corresponding
to $l$ = 102\gra 50\arcmin\ from $b$ = 0\gra 00\arcmin\ to --1\gra 30\arcmin\
from the image in the upper panel. $T_b$ and $N(HI)$ are shown on left and
right y-axes, respectively. The image  shows that the \hi\ brightness temperature is 
$T_b\simeq$ 85 K at $b \simeq$ --0\gra 15\arcmin, while 65\% this value is 
detected at $b \simeq$ --1\gra 30\arcmin.  The {\bf fitted curve}, showed as a 
dashed line, confirms the presence of a $T_b$-gradient.    

The fitting curve was used to estimate the contribution of the \hi\ gas
to the visual absorption $A_v$. 
We calculated the visual absorption $A_v$ due to the molecular and neutral 
hydrogen related to the \hii\ region by using the relation 
$(N(HI) + 2N(H_2))/E(B-V) = 5.8\times10^{21}$ atoms cm$^{-2}$ mag$^{-1}$ 
(Bohlin et al. 1978), where $N(HI)$ and $N(H_2)$ are the \hi\ and H$_2$ 
column densities, respectively, while $E(B-V)$ is the color excess. 
We adopted $E(B-V) = A_v/3.05$ and read off $N(HI)(b)$ from the 
diagram. The estimated \hi\ column densities, as well as the derived
$A_v$-values for each molecular cloud are
included in Table~\ref{tabla_co}.

\section{Photodissociated regions}

The presence of PAH emission at the interfase between the ionized and molecular materia, e.i. at the border of many \hii\ regions, suggests that molecular gas is being photodissociated
by the UV photons emitted by the massive stars. In the following subsections we analyze the presence of PDRs in two selected areas within the \hii\ region, one around WR\,153ab and BD+55\gra 2722,
linked to shell B, and the other around LS+55\gra 39, linked to shell A.

\subsection{Region near WR\,153ab and BD+55\gra 2722}

The upper panel of Figure 10 displays an overlay of the DSS-R image and 
the CO emission 
distribution for a small region of Sh2-132 near WR\,153ab and BD+55\gra 2722.
Thin contours correspond to Cloud A with velocities in the range 
--38.6 to --46.0 \kms, while thick ones delineate Clouds E and F having 
velocities between --46.0 and --53.4 \kms.
The crosses mark the position of the stars. 

The morphological agreement between the optical emission and the 
inner borders of the molecular clouds is excellent, particularly for the 
4\arcmin\  optical filament at $(l,b) \sim$ (102\gra 46\arcmin 30\arcsec,
--0\gra 43\arcmin), where the contour corresponding to 2.25 K closely
follows the optical emission. Note also that difusse emission is
also encircled by molecular emission. The existence of emission at 
8.3 $\mu$m bordering the ionized region (see Fig. 7) strongly indicates 
the presence of a PDR. 

To confirm the existence of a PDR in this region we compared the 
distribution of the ionized and molecular gas, and that of PAHs. 
After convolving the DSS\,R and MSX band A images to the CO beam (46\arcsec),
we averaged the optical,
molecular and band A emissions in concentric rings from the position of the 
WR star within an angle of 300\gra\ from  $(l,b) \sim$ 
(102\gra 52\arcmin,--0\gra 38\arcmin) clockwise up to  $(l,b) \sim$ 
(102\gra 49\arcmin,--0\gra 33\arcmin). The result of this average is displayed 
in the bottom panel of the figure, where intensities in the different bands 
are normalized to the maximum value of each band. The values in the abscisae
indicates distance from the WR star. The diagram reveals a clear 
stratification, with the optical emission peaking near the stellar position,
and the emission at 8.3 $\mu$m peaking  3\arcmin\ far from the star.
The molecular emission increases up to 6\arcmin\ from the star, where the
maximum is present. The observed distribution is typical of molecular clouds
exposed to UV radiation, which photodissociates them (van der Werf et al. 
1996).

\begin{figure}
\centering
\includegraphics[width=84mm]{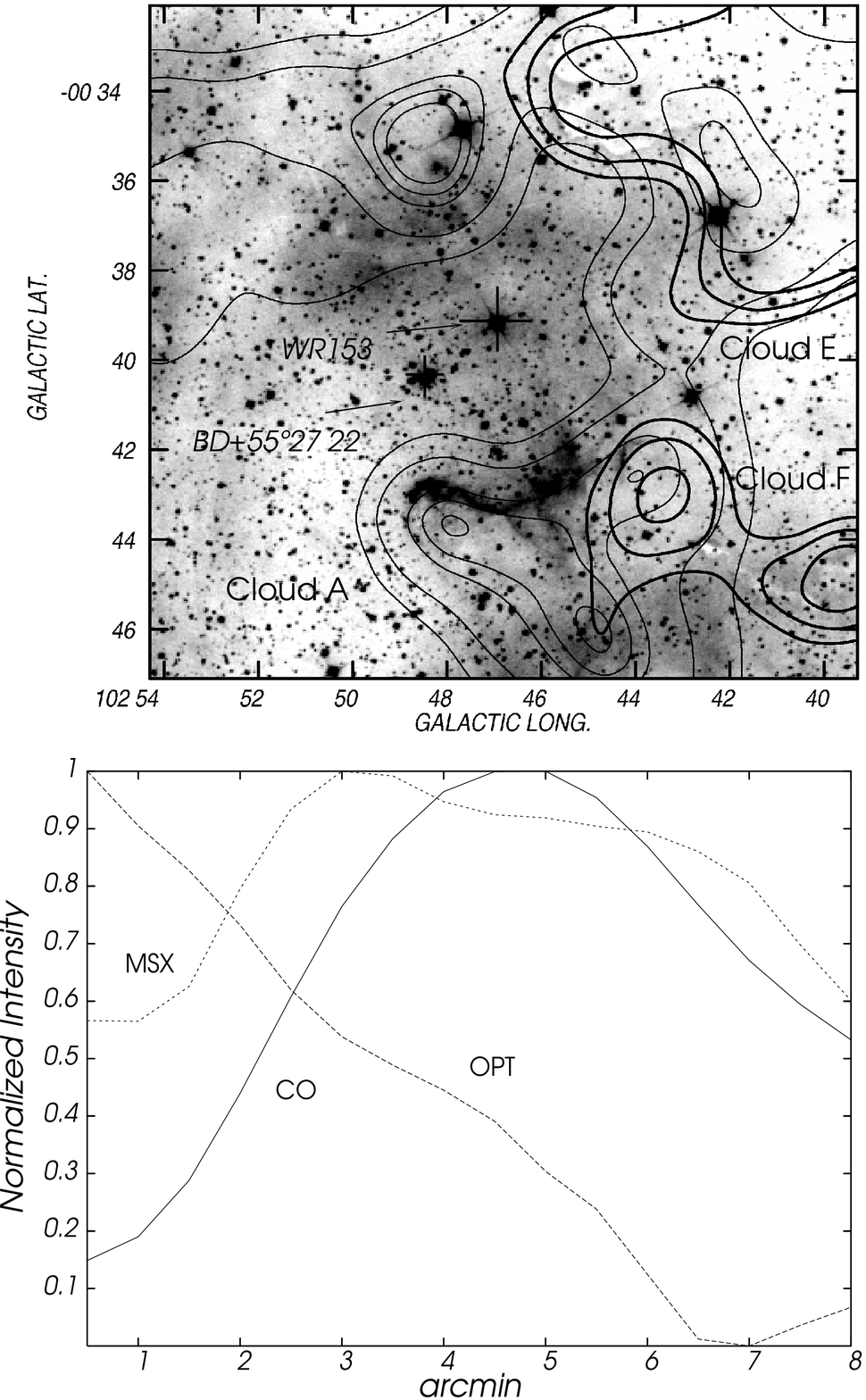}
\caption{{\it Upper panel:} Overlay of the DSS-R image and the CO emission 
in the region near WR\,153 and LS+55\gra 2722, indicated as crosses. 
Thin lines delineate Cloud A and thick lines, clouds E and F.  {\it Bottom panel:} Normalized intensities of the optical, MSX band A, and 
CO emissions in concentric rings as a function of the distance to the WR star.
}
\label{BD}
\end{figure}

\subsection{Region near LS+55\gra 39 }

The region around LS+55\gra 39 is particularly interesting. The upper
panel of Fig. 11 shows that the star is close to a bright optical filament 
of $\sim$2\arcmin\ in size. The fact that the optical filament partially 
follows the border of cloud A indicates that the UV photons of the star are
photodissociating and ionizing the dense gas. 
The high extinction region at $(l,b) \sim$ 
(102\gra 57\arcmin,--0\gra 40\arcmin ) coincides with a section of cloud D. 

With the aim of investigating the presence of a PDR near the optical 
filament, we plotted the normalized intensity of optical, MSX band A, and
CO emissions as a function of the distance to  LS+55\gra 39. The plotted
emission corresponds to the region indicated by the black arrow in the 
upper panel. As for the region around WR\,153ab and BD+55\gra 2722, the 
optical emission peaks closer to the excitation source than the emission in 
MSX band A and the molecular emission, revealing the presence of a 
PDR. 

Taking into account that LS+55\gra 39  is a B0V star with $\log Q(H^0)$ = 
46.23 (Vacca et al. 1996), and assuming that it is immersed in an ionized
interstellar medium  with a mean electron density $n_e$ = 20 cm$^{-3}$, we
can estimate the Str\"omgren radius $r_s$ of the \hii\ region created by
the star from $\frac{3Q(H^0)}{4\pi \alpha_B n_e ^2} = r_s^3$ (Mezger \& 
Henderson 1967), where $\alpha_B$ = 2$\times$10$^{-13}$ \cm s$^{-1}$. 
The result is $r_s$ = 5.3 pc. Bearing in mind that the angular 
distance from the star to the optical filament is 2.2\arcmin, or 
2.25$\pm$0.7 pc at 3.5$\pm$1.0 kpc, we can conclude that the B-type star
is responsible for ionizing the gas in the region and contributing 
to the photodissociation the the molecular gas.

\begin{figure}
\centering
\includegraphics[width=84mm]{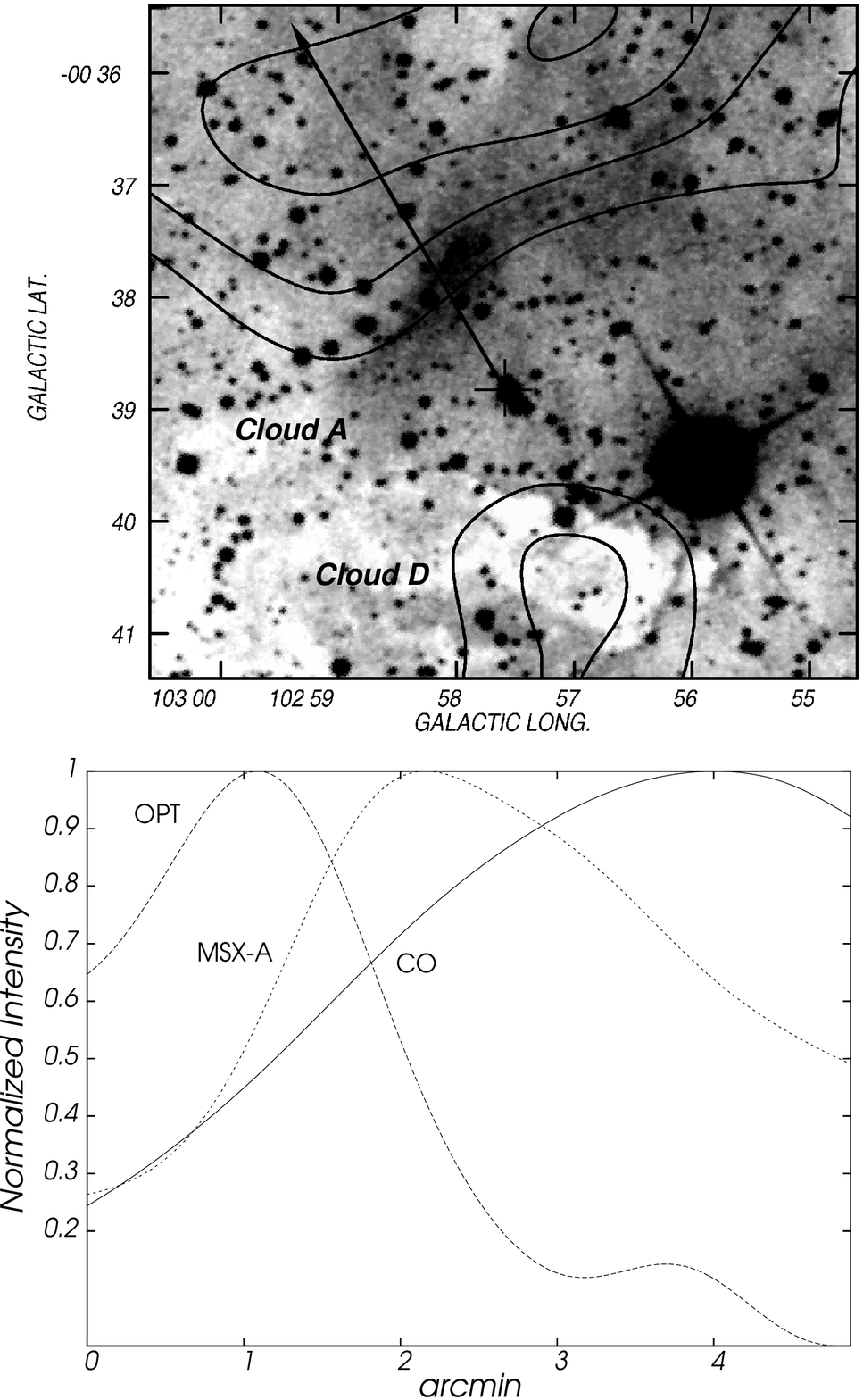}
\caption{{\it Upper panel:} Overlay of the DSS-R image and the CO emission 
in the region near LS+55\gra 39, indicated by a cross. 
Thin lines delineate Clouds A and D. {\it Bottom panel:} Normalized intensities of the optical, MSX band A, and CO emissions in concentric rings as a function of the distance to the B star.}
\label{LS}
\end{figure}

\section{Scenario and conclusions}

In this paper we have analyzed the distribution of the ionized, neutral
atomic and molecular gas and that of the interstellar dust in 
Sh2-132. 

This study revealed the presence of a number of molecular clouds 
encircling the ionized nebula, with a mass of 9000 \Msun. 
The velocity of the molecular gas, which is
in the range from --38.6 to --53.4 \kms\ coincides with the velocities of 
the $H\alpha$ line, and $He$ and $H$ radio 
recombination lines, strongly indicating that the ionized gas is associated
with the nebula and interacting with the ionized gas. 

The ring-like appearance of the gas and dust distribution in the
environs of the WR star suggests the action of stellar winds, that
sweep up and compress the gas. 
The mechanical energy $E_\mathrm{w}$ released by the multiple system WR\,153ab into the ISM can be roughly estimated taking into account 
the two WR components of the binary system during the duration of the WR 
phase of a massive star, the previous main sequence phase of the current 
WR stars, and the two O-star components. We adopt a mass loss rate 
$\log$ ($\dot{M}$) = --6.64 and a terminal velocity $v_{\infty}$ = 1950 \kms, 
corresponding to a O8.5V star (Smith et al. 2002) as parameters of the
stellar wind of the O-type components. For the WR components we adopt 
a conservative mass loss rate 
$\dot{M} = $5$\times$10$^{-6}$\,M$_\odot$\,yr$^{-1}$ and a terminal
velocity $v_{\infty}$ = 2000~\kms. Adopting lifetimes of 3$\times$10$^6$ yr 
and 0.5$\times$10$^6$ yr for the O and WR phases of the stars, 
respectively (Conti \& Vacca 1990, Meynet \& Maeder 2005), 
$E_\mathrm{w}$ results to be $\sim$ 
7.2$\times$10$^{50}$ erg. 

  Considering the molecular clouds E, F and a section of cloud A, and the dust  and ionized components, with an expansion velocity of 7.4 \kms, the kinetic energy of shell B is $\sim$2.80$\times$10$^{48}$ erg. This result indicates that the multiple system can shape shell B.

The emission at 8.3 $\mu$m appears concentrated in a ring-like structure
bordering the brightest sections of the nebula, indicating that 
molecular material is being photodissociated at the interface between the 
ionized and molecular gas. This configuration resembles the IR bubbles
analyzed by Watson et al. (2008). To confirm the presence of photodissociation 
regions we have analyzed in some detail two regions within the complex. 
One in Shell B around WR\,153ab and BD+55\gra 2722, and the other in 
Shell A around LS+55\gra39. The relative positions of the ionized gas, 
PAHs, and molecular gas in these regions show the stratification typical 
in PDRs. Thus, the distribution of $H_2$ plays a main role in shaping the 
\hii\ region.

The emission distribution at 60 $\mu$m shows the same ring-like 
structure, along with emission inside the ring. The presence of emission
in the far IR inside the \hii\ region, which originates from large dust 
grains, is compatible with the
existence of PDRs since PAHs are destroyed inside \hii\ regions, while
large grains are not. 

 The number of UV photons emitted by the massive stars
in the region is seven times higher than the UV photons used to ionize the gas,
indicating  that the massive stars are responsible for the ionization
of the region. Clearly, a large fraction of the available UV photons
are used in the heating of the interstellar gas and in the photodissociation 
of the dense material.

  Different studies of \hii\ regions such as those towards RCW\,79 (Zavagno et al. 2006), RCW\,82 (Pomar\'es et al. 2009), and Sh2-212 (Deharveng et al 2008) show how the interaction between the massive stars with the surrounding molecular material induces star forming regions through the ``collect and collapse'' process (Elmegreen and Lada 1977) in the borders of these \hii\ regions. For example, the paper on RCW\,79 (Zavagno et al. 2006) reveals the formation of YSOs inside pre-existing molecular cores located at the border of the ionized region. By using the analytical model by Whitworth et al. (1994) to determine the time at which the fragmentation occurs and the size of the structure at that time, they conclude that the collect and collapse process may not be the only one at work. Sh2-212 (Deharveng et al. 2008) and Sh2-104 (Deharveng et al. 2003) represent good examples of stellar formation in which the molecular clouds surround the \hii\ regions and the PAH emission is located at the interfase between the ionized and the molecular gases. In particular, Sh2-104 is encircled by a molecular ring showing four dense molecular clouds containing many dense cores. This configuration is a strong evidence in favour of the triggering star formation process descripted by the ``collect and collapse'' model. In all of the preceding examples, we have an \hii\ region generated by an O-type star or a cluster of massive stars. 

  The region around Shell B presents a similar interstellar scenario as the preceding examples, e.i an optical shell bounded by a strong arc-like feature of PAH emission which are surrounding by molecular material (Clouds A, E, and F). In such scenario, star formation can be induced by the action
of the massive stars through the ionizing flux and stellar winds. Indeed,
the presence of YSOs towards the molecular clouds linked to Sh2-132
indicates that star formation is very active in this region of the Galaxy
(Vasquez et al. in preparation).

\section*{acknowledgements}

We acknowledge the anonymous referee of her/his comments.
This project was partially financed by the Consejo Nacional de 
Investigaciones Cient\'{\i}ficas y T\'ecnicas (CONICET) of Argentina under 
projects PIP 112-200801-02488 and PIP 112-200801-01299,  Universidad Nacional 
de La Plata (UNLP) under project 11/G093, Universidad de Buenos Aires under 
project UBACyT X482, and Agencia Nacional de Promoci\'on Cient\'{\i}fica y 
Tecnol\'ogica (ANPCYT) under projects PICT 00812 and 2007-00902.
The Digitized Sky Survey (DSS) was produced at the Space Telescope Science
Institute under US Government grant NAGW-2166. This work was partly (S.P.) 
supported by the Natural Sciences
and Engineering Research Council of Canada (NSERC) and the Fonds
FQRNT of Qu\'ebec.  The  DRAO Synthesis
Telescope is operated as a national facility by the National Research 
Council of Canada. The CGPS is a Canadian project with
international partners and is supported by grants from NSERC.
Data from the CGPS
is publicly available through the facilities of the Canadian
Astronomy Data Centre (http://cadc.hia.nrc.ca) operated by the
Herzberg Institute of Astrophysics, NRC.

\end{document}